\documentclass{desyproc}

\begin{document}
\title{Searching for Axion Dark Matter using Radio Telescopes}

\author{{\slshape Katharine Kelley$^1$, P. J. Quinn$^{1}$}\\[1ex]
$^1$International Centre for Radio Astronomy Research (ICRAR), University of Western Australia, Ken and Julie Michael Building, 7 Fairway, Crawley, WA 6009, Australia}

\contribID{familyname\_firstname}

\confID{13889}  
\desyproc{DESY-PROC-2017-XX}
\acronym{Patras 2017} 
\doi  

\maketitle

\begin{abstract}

We investigate the use of next generation radio telescopes such as the Square Kilometre Array (SKA) to detect axion two-photon coupling in the astrophysical environment.  The uncertainty surrounding astrophysical magnetic fields presents new challenges, but with a frequency range corresponding to axions of mass $1.7-57\mu$eV and a spectral profile with a number of distinguishing features, SKA-mid offers a tantalising opportunity to constrain axion dark matter properties.  To determine the sensitivity of SKA-mid to an axion signal, we consider observations of the Galactic centre and interstellar medium, and find that this new telescope could allow us to probe axion couplings $\gtrsim10^{-16}$GeV$^{-1}$.

\end{abstract}

\section{Introduction}\label{intro}

In this paper we motivate the use of radio telescopes in the search for cold dark matter (CDM) axions.  We focus on the axion predicted by Weinberg~\cite{1978PhRvL..40..223W} and Wilczek~\cite{1978PhRvL..40..279W} following the introduction of a new $U(1)$ symmetry by Peccei \& Quinn~\cite{1977PhRvL..38.1440P}, and assume that it comprises all dark matter in the halo of our Galaxy.  This axion has a mass constrained to be in the range $1-1,000\mu$eVc$^{-2}$~\cite{1978PhRvD..18.1829D, 1980PhRvD..22..839D, 1987PhRvD..36.2211R,1989PhRvD..39.1020B,1983PhLB..120..127P, 1983PhLB..120..137D}, for which conversion in a static magnetic field would produce a spectral line in the frequency range $0.2-200$GHz.  It is this property that leads us to consider whether next generation radio telescopes such as the Square Kilometre Array (SKA) may contribute to detection efforts.

\begin{wrapfigure}{r}{0.5\textwidth}
\begin{center}
\includegraphics[width=0.48\textwidth]{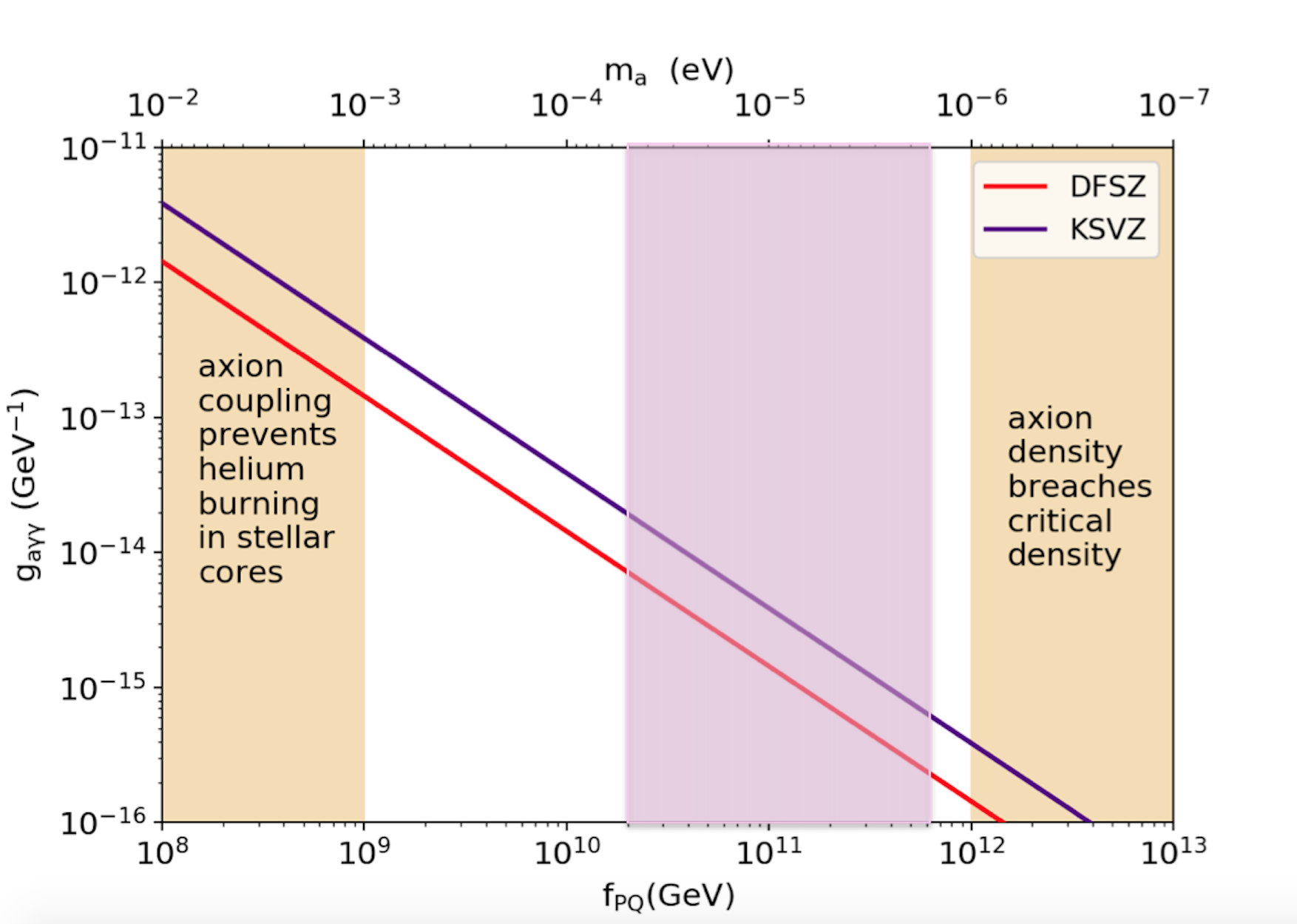}
\end{center}
\caption{The mass, $m_a$, and coupling strength, $g_{a\gamma\gamma}$, of the axion are determined by the P-Q symmetry breaking scale, $f_{pq}$, and constrained by astrophysical observations. The reference to KSVZ and DFSZ relates to the two generally accepted axion models from which its properties are derived, and the pink shaded area shows the breadth of parameter space that could be probed based on the frequency range of SKA-mid.}
\label{coupling}
\end{wrapfigure}

The SKA is a radio telescope planned for construction in Australia and Southern Africa.  When built, it will be the largest radio telescope in the world and will provide sensitivity and spatial resolution at least an order of magnitude better than current technologies.  The array comprises two key components, SKA-low with frequency range $100-350$MHz, and SKA-mid with frequency range $0.4-13.8$GHz.  For axion conversion in the dark halo of our Galaxy, SKA-mid's construction as an array of parabolic dishes offers the best opportunity for detection, its frequency range corresponding to an axion mass of $1.7-57\mu$eV (see Figure \ref{coupling}).  The timeline for construction is approximately $10-15$ years, however there exist today two precursor telescopes, the Australian SKA Pathfinder (ASKAP) and the Karoo Array Telescope (MeerKAT), which are already providing new and valuable data.  To investigate whether these radio telescopes could indeed contribute to the global search for axion dark matter, we set out below in Section \ref{sec: coupling} our key considerations for axion conversion in an astrophysical magnetic field, and assess in Section \ref{sec: flux} the coupling strength that may be probed by observing the Galactic centre and interstellar medium.

\section{Axion two-photon coupling in astrophysical magnetic fields}\label{sec: coupling}

The photon production rate associated with axion conversion in a magnetic field is well documented in the literature (most notably~\cite{1983PhRvL..51.1415S}).  We revisit this calculation with the astrophysical environment in mind and find a few key areas for consideration: (1) The spatial profile of the magnetic field plays an important role in determining both the photon production rate and the properties of the real photon produced; (2) The astrophysical environment will contain non-static as well as static fields, and the production of higher energy photons in a strong non-static field is worthy of consideration; and (3) The frame of axion conversion relative to the observer's frame must be taken into consideration when establishing the width of the spectral profile.

We find that in considering the non-relativistic cold dark matter axion, the magnetic field is required to provide momentum to the interaction, and as a result only certain modes of the magnetic field contribute to the photon production rate.  This is required by energy/momentum conservation and is one of the principal differences with laboratory based experiments seeking to detect the same phenomenon.  With the momentum of the axion, ${\bf k}_a$, assumed to be negligible, only those modes which satisfy $\Delta E =m_ac^2$ will contribute to the interaction, where $\Delta E=\hbar k_{\gamma'} c - \hbar \omega_{\gamma'}$, $\hbar \omega_{\gamma'}$ is the energy of the virtual photon and $\hbar k_a$ its momentum.  In the case of a static field for which $\hbar \omega_{\gamma'}=0$, this reduces to those modes with $k_{\gamma'}=m_ac/\hbar$. The photon production rate, $\Gamma(N_a)$, for axion conversion in a non-static inhomogeneous field can then be written as,

\begin{eqnarray}\label{nonstatic}
\Gamma(N_a)\approx 2.2 \times 10^{25} \textrm{Tesla$^{-2}$kg$^{-1}$ms$^{-2}$}\ \left(\frac{\rho_{DM}}{T}\right)\ \int{d^3{\bf k}_{\gamma}}\ k_{\gamma}\left| B({\bf k}_{\gamma},\omega_{\gamma}-\omega_a) \right|^2  \nonumber
\end{eqnarray}

\noindent where $T$ is the time over which the temporal fluctuation of the magnetic field repeats, $\rho_{DM}$ is the dark matter density, and $B({\bf k}_{\gamma},\omega_{\gamma}-\omega_a)$ is the Fourier transform of the classical magnetic field at a given energy and momentum.  For a static-field this photon production rate can then be simplified to,

\begin{eqnarray}\label{static}
\Gamma(N_a)\approx 1.3 \times 10^{20} \textrm{Tesla$^{-2}$kg$^{-1}$s$^{-1}$}\left(\frac{\rho_{DM}}{V}\right) \left| B\left(\frac{m_ac}{\hbar}\right) \right|^2 
\end{eqnarray}

\noindent where $V$ is the volume observed.  Important assumptions must be made about the astrophysical environment in order to apply these equations.  The magnetic field must be inhomogeneous on scales that can be seen by the axion, and the plasma density must be less than that required for resonance conversion.  While there are great uncertainties in the small scale structure of the Galaxy, the average electron density of order $1$cm$^{-3}$ and the extremely high plasma density required for resonance conversion of a non-relativistic axion, leads us to believe that there is merit in investigating the use of radio telescopes further.

\section{Observing axion conversion in the interstellar medium with the SKA}\label{sec: flux}

\begin{wrapfigure}{r}{0.5\textwidth}
\begin{center}
\includegraphics[width=0.48\textwidth]{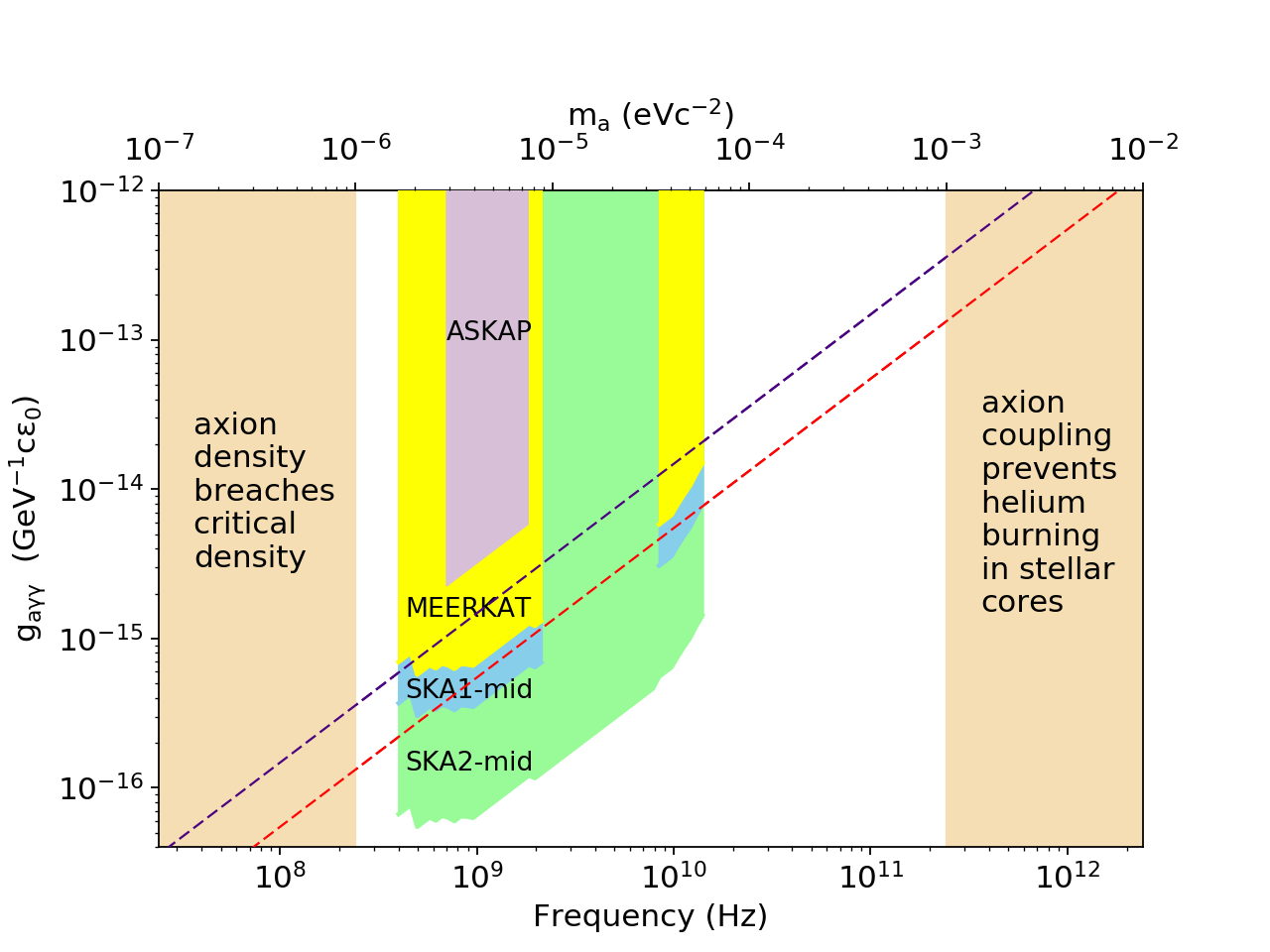}
\end{center}
\caption{The sensitivity of SKA-mid shows considerable improvement on the pre-cursor telescopes, the Australian SKA Pathfinder (ASKAP) and the Karoo Array Telescope (MeerKAT).  In this Figure we show the coupling strength that could be probed by observing the interstellar medium across the frequency range accessible to ASKAP, MeerKAT and SKA-mid.  The system temperature of the SKA is minimised between $\sim 2 - 7$GHz, corresponding to an axion mass of $\sim 8.26 - 28.91\mu$eVc$^{-2}$ and providing a good opportunity for detection of both the KSVZ and DFSZ axion. Figure taken from~\cite{2017ApJ...845L...4K}.}
\label{sensitivity}
\end{wrapfigure}

In order to determine the energy produced in a given region in the Milky Way, we must make an assumption about both the density of dark matter at a given point and the strength and profile of the magnetic field.  We use Equation \ref{static} to determine the photon production rate at the Galactic centre and throughout the interstellar medium, making the assumption that the related magnetic field does not change on timescales visible to the axion.  

We calculate the density of dark matter using an adjusted NFW profile with a central core density of $10$GeVcm$^{-3}$ and a density at Earth of $0.4$GeVcm$^{-3}$.  The strength of the magnetic field is assumed to have a strength of $50\mu$G within a radius of 1kpc of the Galactic centre and a profile that reduces radially along the Galactic disk as $r(kpc)^{-1}$.  Within each region we then assume that the profile of the magnetic field is turbulent on small scales and can be described by a Kolmogorov spectrum.  

To calculate the flux at Earth that results from photons being produced via axion conversion across the interstellar medium, we assume that the energy propagates in all directions and is shared across the surface of a sphere, with radius given by the distance between the region of conversion and the Earth.  We also assume that the dark halo of the Milky Way displays no net rotation, and that the Galactic disk is moving through this halo with a rotational velocity at Earth of 220kms$^{-1}$.  Using these assumptions for an axion of mass $2.05\mu$eV the flux at Earth is calculated to be $3.7\mu$Jy.  The axion signature will be observed as a spectral line with a central frequency of $\sim500$MHz and a width given by the axion velocity and relative frames of the conversion and the observer, in this case $\sim 200$kHz.  The polarisation of the signal should also trace the polarisation of the magnetic field when observing a coherent single source.  Such distinguishing features should allow an axion line to be identified from astrophysical foregrounds and other spectral lines.  In particular, synchrotron radiation which makes up a large portion of foregrounds at this frequency should have a polarisation perpendicular to the axion signal.

Using this flux, we are able to compare the strength of the axion signal with the limiting sensitivity of SKA-mid.  Using only a 24 hour integration time we find that the limiting sensitivity of SKA-mid is 0.04mK, compared to a signal temperature of 1.17mK when utilising the full (1km)$^2$ collecting area of SKA-mid Phase 2 to observe axion conversion in the Galactic centre and interstellar medium.  Figure \ref{sensitivity} shows the coupling strength that could be probed by such observations, and it is clear that this technology has the potential to open up a new window for the detection of axion dark matter.

\section{Acknowledgments}

We would like to thank the conference organisers for providing a forum for fruitful discussion.  We also thank Prof. Ian McArthur at the University of Western Australia for his invaluable guidance in preparing this work.  

\section{Bibliography}


\begin{footnotesize}

\end{footnotesize}


\end{document}